# Muon Telescope at BEO – Moussala *

**Ivo Angelov [1], Elisaveta Malamova [2], Jordan Stamenov [2]**

[1]South West University "N. Rilski" ( Bulgaria ) , address for correspondence : i_angeloff@mail.bg
[2] Institute For Nuclear Research and Nuclear Energy – Bulgarian Academy of Science

## Abstract

A muon telescope based on 8 water cherenkov detectors was mounted at BEO Moussala – Institute for Nuclear Research and Nuclear Energy – Bulgarian Academy of Science. The telescope will be used for continuous measuring the intensity of the muon component of the cosmic rays, exploring its variations and possible correlations with environment parameters. A detailed description of the device is presented in this paper.

## 1. Description of the telescope.

### 1.1. The Detectors

The detectors of the telescope are of the same type as the detectors used in the muon telescope at the Department of Physics – South West University "N. Rilski". [1]

Each of the detectors includes a mirror tank with distilled water and photomultiplier with preamplifier mounted at its housing. (fig. 1. ) The dimensions of the detectors are 50x50x12.5 cm, and the distilled water layer used as radiator is 10 cm. When a cosmic rays muon passes through the radiator, cherenkov ligh is generated if the energy of the muon is high enough that its speed is greater than the speed of the light in the water. Part of the cherenkov photons reach the photocathode of the photomultiplier after multiple reflections from the mirror walls of the container. The photomultiplier converts this light pulse into an electric pulse, which is amplified by the build-in preamplifier and then sent through a coaxial cable to the next stages of the telescopes electronics.

The 2.5" photomultiplier tubes type FEU-110 or FEU-139 are used. The preamplifier is a low noise one and has amplification x50 and 90 Mhz bandwidth.

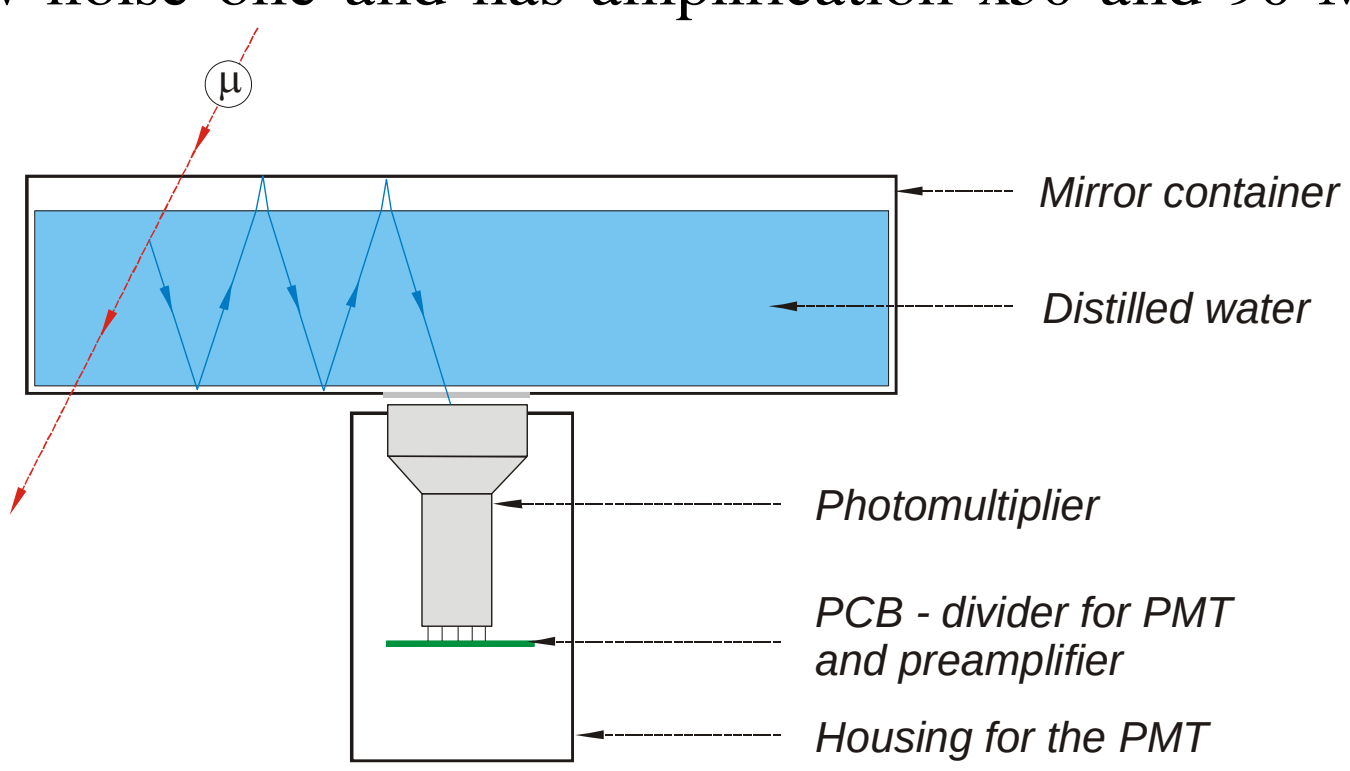

*fig. 1. A water cherenkov detector*

* Preprint submitted to "Scientific Research" - a electronic issue of South West University "N. Rilski" ( Bulgaria )
http://press.swu.bg/epub/

### 1.2 The Detectors Setup

The detectors are mounted in two parallel planes, 4 in each. (fig. 2. ) The distance between the detectors planes is 1m. The signals from the detectors are connected to coincidence circuits, this way counting only cherenkov pulses from muons passing through pair of detectors – one from the upper plane and other from the down - plane is provided. 5 different angular intervals are determined using combinations of pairs of detectors connected to coincidence circuits. The outputs of the coincidence circuits are connected to counters which count the number of coincidences from each pair of detectors for 1 minute intervals. (In fact this is the number of cosmic rays muons, passing through the two detectors.)

A 8 cm thick lead absorber layer is mounted between the two detector planes to reject the electron component of the cosmic rays.

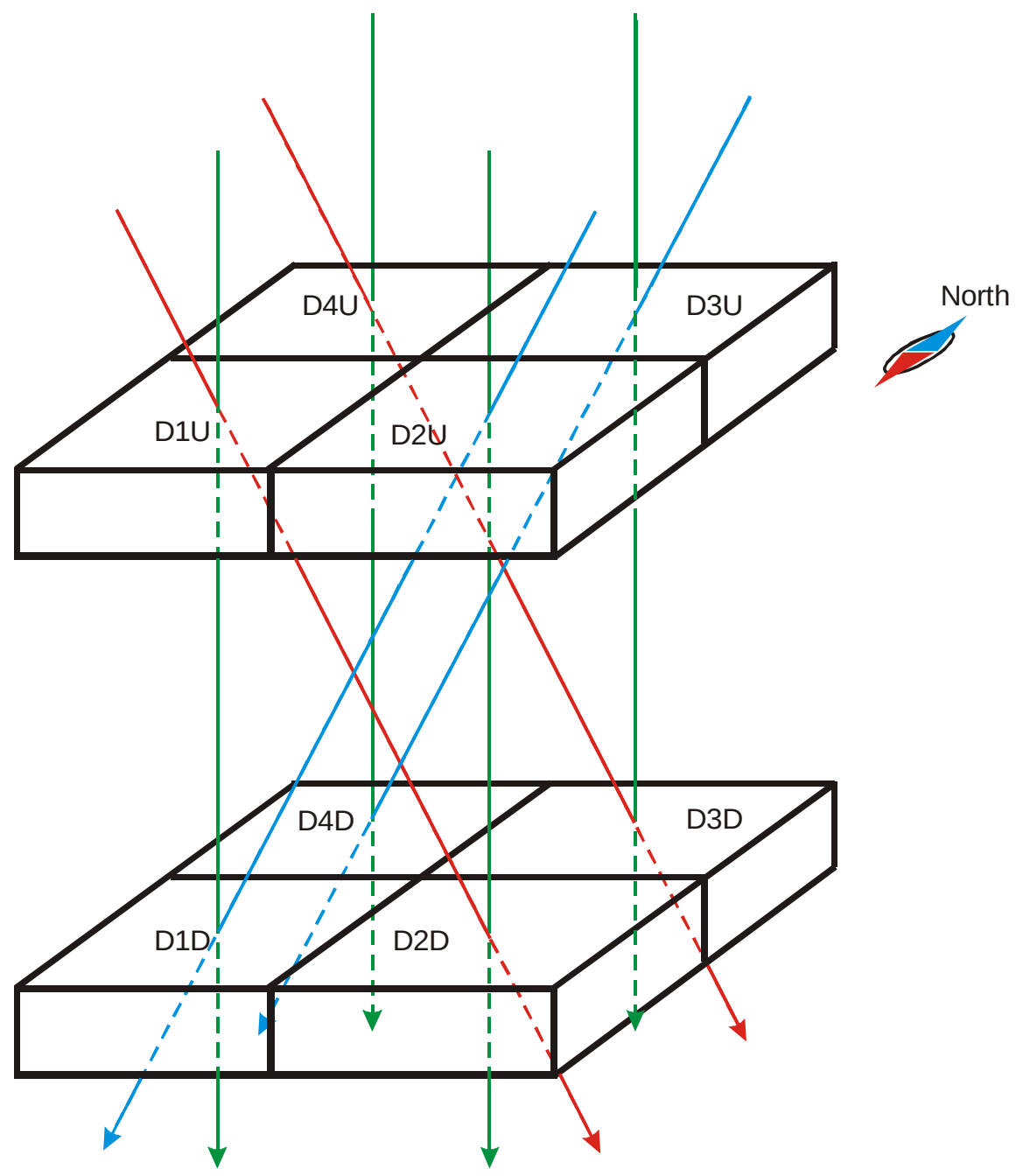

*fig.2 Placement of the detectos of the telescope.*

| Direction | Zenith angular interval | Number of detector pairs | Detectors Connected to Coincidence circuits | Counters |
|---|---|---|---|---|
| Vertical | +25.6° -25.6° | 4 | D1U-D1D, D2U-D2D, D3U-D3D, D4U-D4D | C1, C2, C3, C4 |
| South-North | 0 - 45° | 2 | D1U-D4D, D2U-D3D | C5, C6 |
| North-South | 0 - 45° | 2 | D4U-D1D, D3U-D2D | C9, C10 |
| West-East | 0 - 45° | 2 | D1U-D2D, D4U-D3D | C7, C8 |
| East-West | 0 - 45° | 2 | D3U-D4D, D2U-D1D | C11, C12 |

*Table 1. Coincidence circuits and angular intervals*

### 1.3 The registration part

The block schematic of the registration part is shown at fig.3

Each of the signals from the detectors is connected through a coaxial cable to a discriminator. The discriminators form 60 nanoseconds digital pulses if the incoming pulse exceeds the given for the discriminator threshold. The threshold can be regulated in the range 15-50 mV. ( The actual threshold adjusted is 28 mV.) The formed pulses are led to the inputs of the coincidence circuits.

The coincidence circuits consist of 12 fast AND logical elements, connected as drawn in the schematic. A pulse is formed at the output of a coincidence circuit if pulses appear at its input simultaneously. (When we have a particle passed both the detectors.) The outputs of the coincidence circuits are connected to the inputs of the counters circuit.

The counters circuit has 12 binary 8-bit counters and one 24 bit counter. The 8 bit counters count the pulses from the coincidence circuits for 15 s time intervals. The time intervals are determined by quartz stabilized timer. The value from each counters register is written in a file on a hard disc drive of the data acquisition personal computer after the time interval is finished, and counting for the next interval is started.

The 24 bit counter counts every time interval the count rate of one of the detectors, and these data is used to control the proper operation of the detectors.

The high voltage power supply provides stabilized voltage with low ripple amplitude to power the photomultiplier tubes. The main high voltage is regulated in the range 1500 – 1950 V. A separate regulated in 15 steps of 25 volts output is present for each of the photomultipliers.

The data acquisition personal computer is a 586 family. The parallel port is used as an interface to the counters circuits. The software is working in any WINDOWS operational system and visualizes on the monitor screen each counter state, a relative plot of the count rate for the different angular components of the muon flux and writes the data in formatted ASCII files on the hard disk drive.

The data are recorded on the hard disk of the data acquisition PC in three types of files. The raw data are the coincidence count rates for 15 seconds time intervals. The averaged data present the average count rate for each direction for 5 minute intervals, and its relative variations. The third type of data recorded is the individual dark count rate of each PMT for 15 second interval, once in 2 minutes.

## 2. Setting in operation and adjustment of the electronics. Main characteristics of the telescope.

As each of the parts of the telescope has passed separate laboratory tests before delivering to the observatory, the first thing to do was to make some complete tests of the whole telescope, checking the oscilograms at the test points of the electronic circuits.

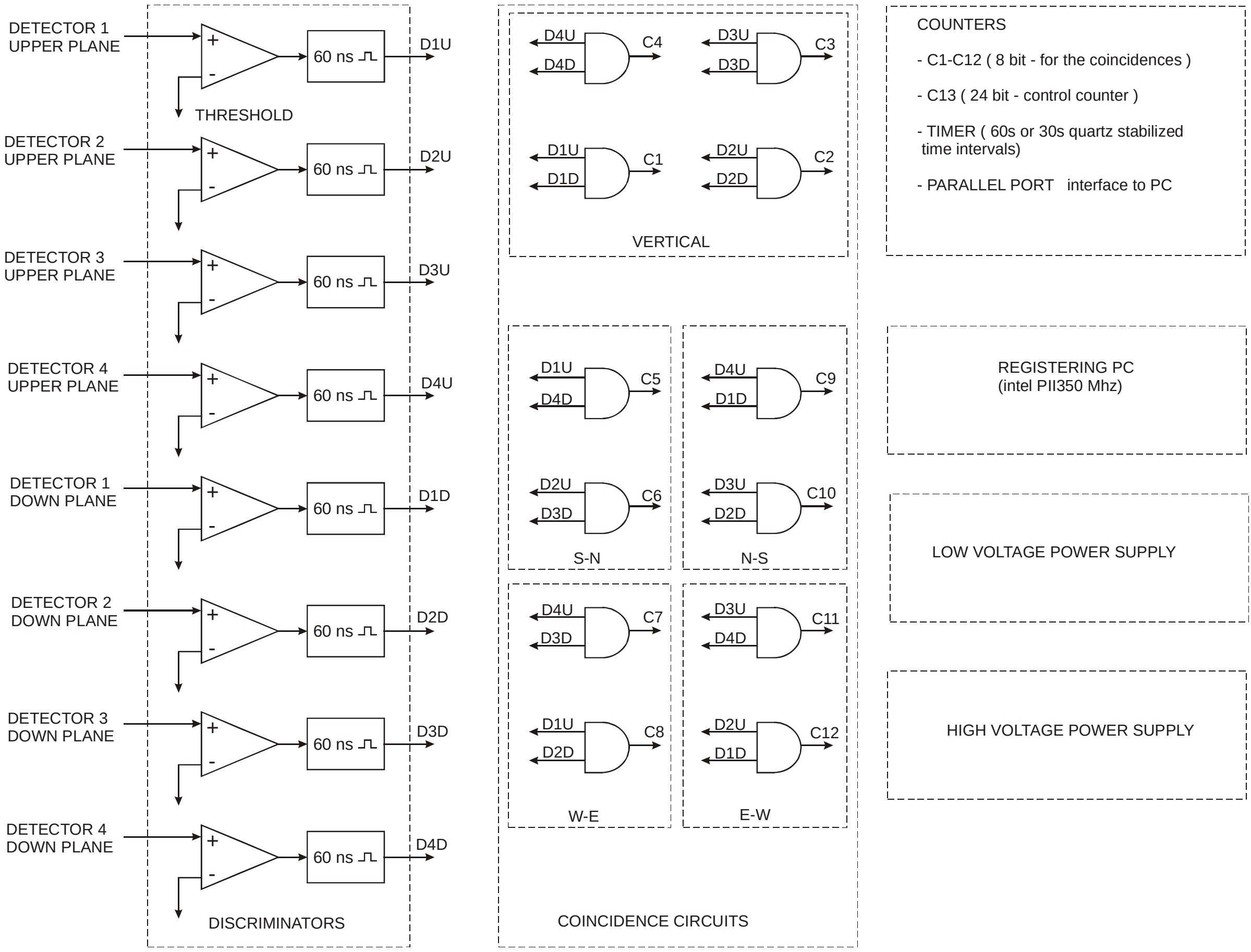

*fig.3 Block-schematic of the telescope registering part.*

After successful tests and check of the dark count rate of each detector, the optimal high voltage and the threshold of each discriminator was experimentally set.

The count rate of each detector is of the order of ~ (3-8).$10^4$ $min^{-1}$ (including the dark counts of the PMT), increasing the high voltage of the PMT increases its gain and makes the detector sensitivity greater , decreasing the threshold level also brings the efficiency of the detector higher. But both these factors rapidly increase the dark counts of the PMT, and the number of the random coincidences, in spite of the short pulse widths, becomes intolerable high. The number of random coincidences should be kept as low as possible, otherwise the statistical error of the telescope will be poored.

Because of specific parameters of each of the PMT, (dark counts, gain are different for each PMT, although they are one and the same type, this is a restriction during their production – there are always tolerances) the optimal high voltage and the discriminator threshold can be determined only by experiment. The dependencies of the single counting rates for each pair of detectors (one with the one above it) and the number of vertical coincidences per minute on the high voltage and on the discriminator threshold over the ranges 1500-1950 V , 15mV–50mV had to be taken and the values for the operation of the given detector to be selected, looking for highest detectors efficiency and lowest random coincidences.

The dependence of the count rate for the detector pairs counting the muon flux in vertical direction on the high voltage power supply of the photomultiplier tubes is presented on figure 4.

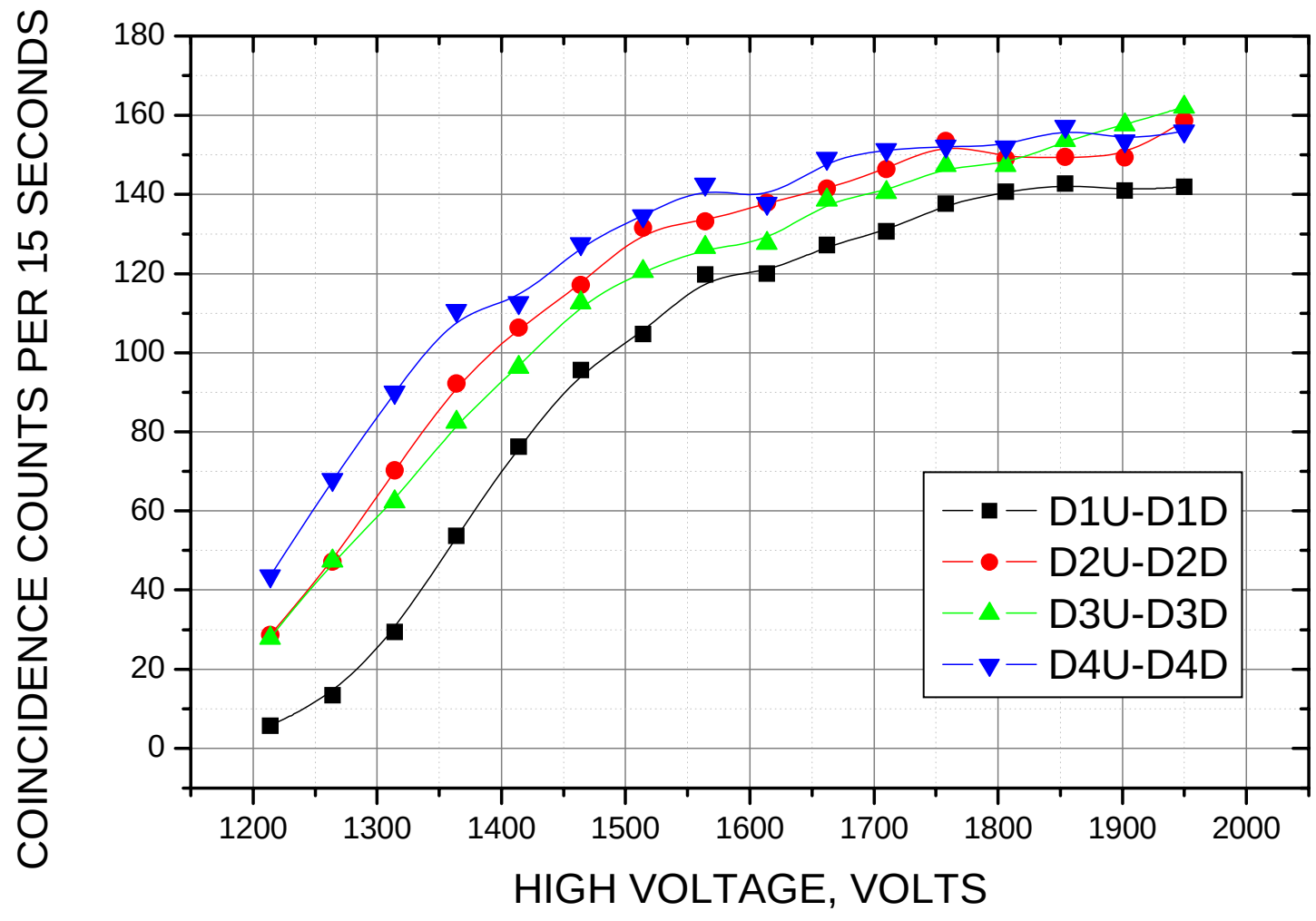

*fig.4 Dependance on the count rate of the detector pairs on the high voltage power supply of the PMT. (discriminator threshold = 28 mV)*

The high voltage for each PMT was selected according this plot, taking in mind that the detectors should work with maximum efficiency, at the plateau (1700 – 1900 V) of the counting characteristic, but with minimum dark count rate. (also available as data and plot )

The energy threshold of the telescope is ≈0.5 GeV, and the count rate for each of the 4 pair of detectors measuring the intensity of the cosmic rays muons in vertical direction is ≈580 $min^{-1}$. The statistical error of the telescope is better than 1% for 5 minutes and 0.4% for ½ hour time intervals and vertical component measurements.

## 3. First results.

The barometric effect for cosmic rays muons is presented on fig.5 . A clear negative correlation is visible, proving the proper operation of the device. Software for automatic pressure correction of the raw data is in process of development now.

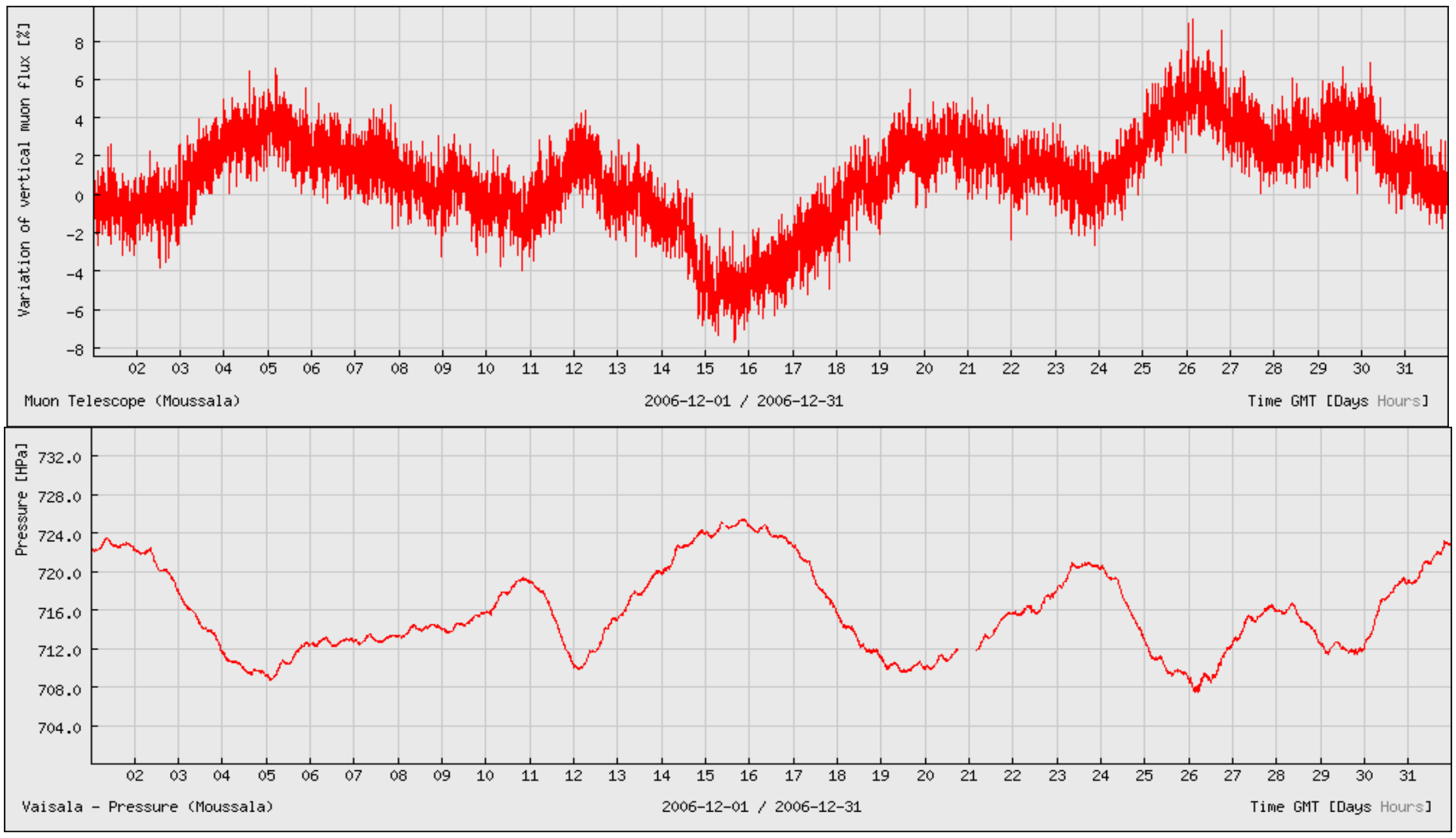

*fig.5 Plots of the measured intensity of the muon flux (vertical component, averaged for 5 minute intervals ) and the atmospheric pressure for December 2006.*

## 4. Conclusion.

The muon telescope is completely mounted and in continuous operation providing data since 1st august 2006. Plots of the data are available on-line at the web-page of the observatory http://beo-db.inrne.bas.bg/moussala/ .

The accuracy of the telescope is better than expected because of the higher muon flux intensity at this altitude, giving higher count rates and lower statistical errors.

## Acknowledgements

The telescope was constructed and fully funded as a part of the **FP6 BEOBAL** project - BEO Centre of Excellence Research Capacity Improvement for Sustainable Environment and Advanced Integration into ERA .

## References

1. E. Malamova, I. Angelov, I. Kalapov, K. Davidkov, J. Stamenov , **Muon Cherenkov Telescope** , Proceedings of 27 ICRC Hamburg 2001, p. 3952-3955
2. Merson, G. I., et all, Water Cerenkov Detector with Focusing , Questions of Nuclear Science and Techniques, 4, 1988
3. Dorman, L. I. , Variations of Cosmic Rays, Publishing House for Technical and Theoretical Literature, Moscow 1957
4. Murzin, V. S. , Physics of Cosmic Rays, Moscow University Publishing House, 1969
5. Dorman, L. I. , Variations of Galactic Cosmic Rays, Moscow University Publishing House, 1975

**Some pictures of the device are available at the last pages of this paper.**

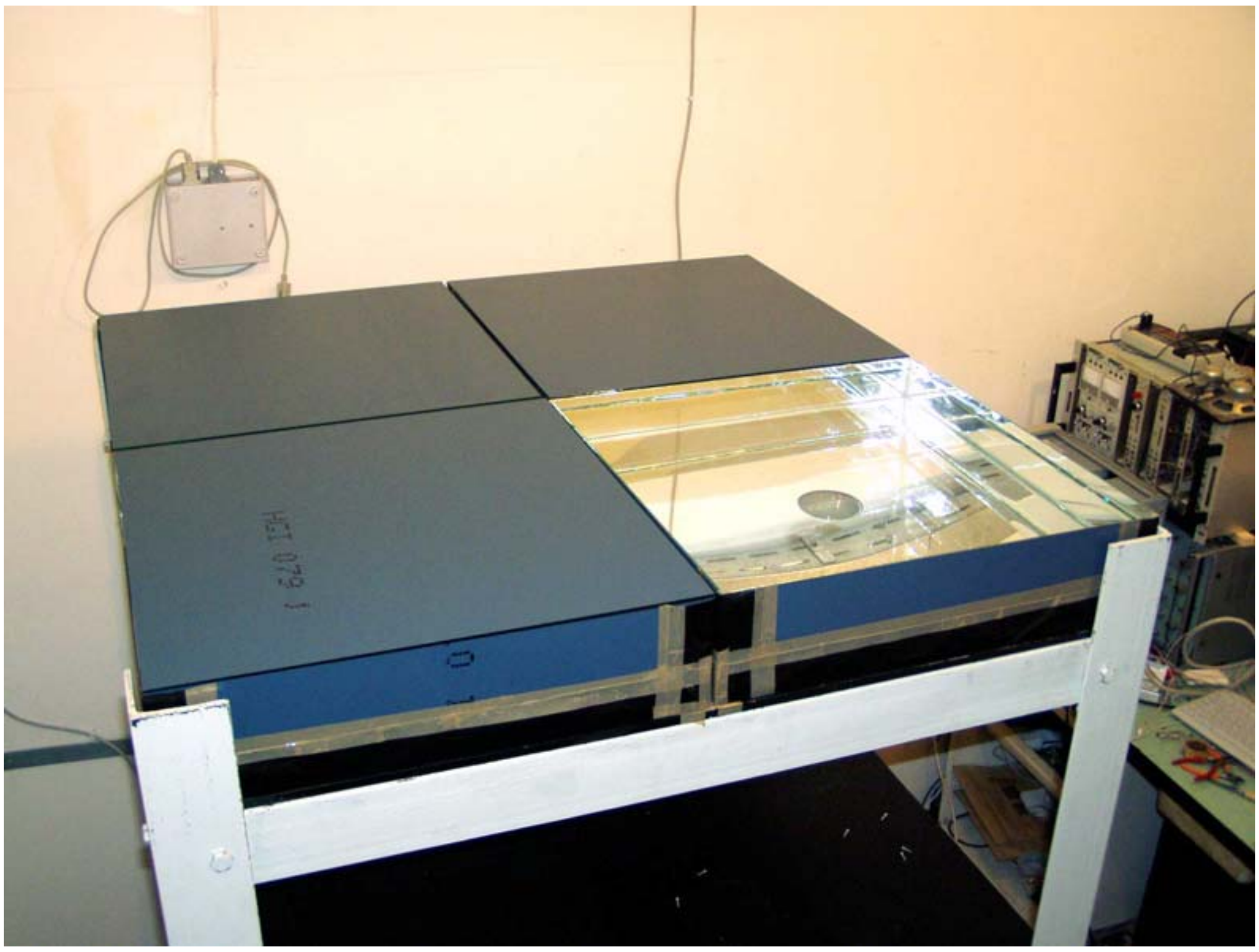

*pic. 1. The cherenkov detectors from the upper plane during assembling.*

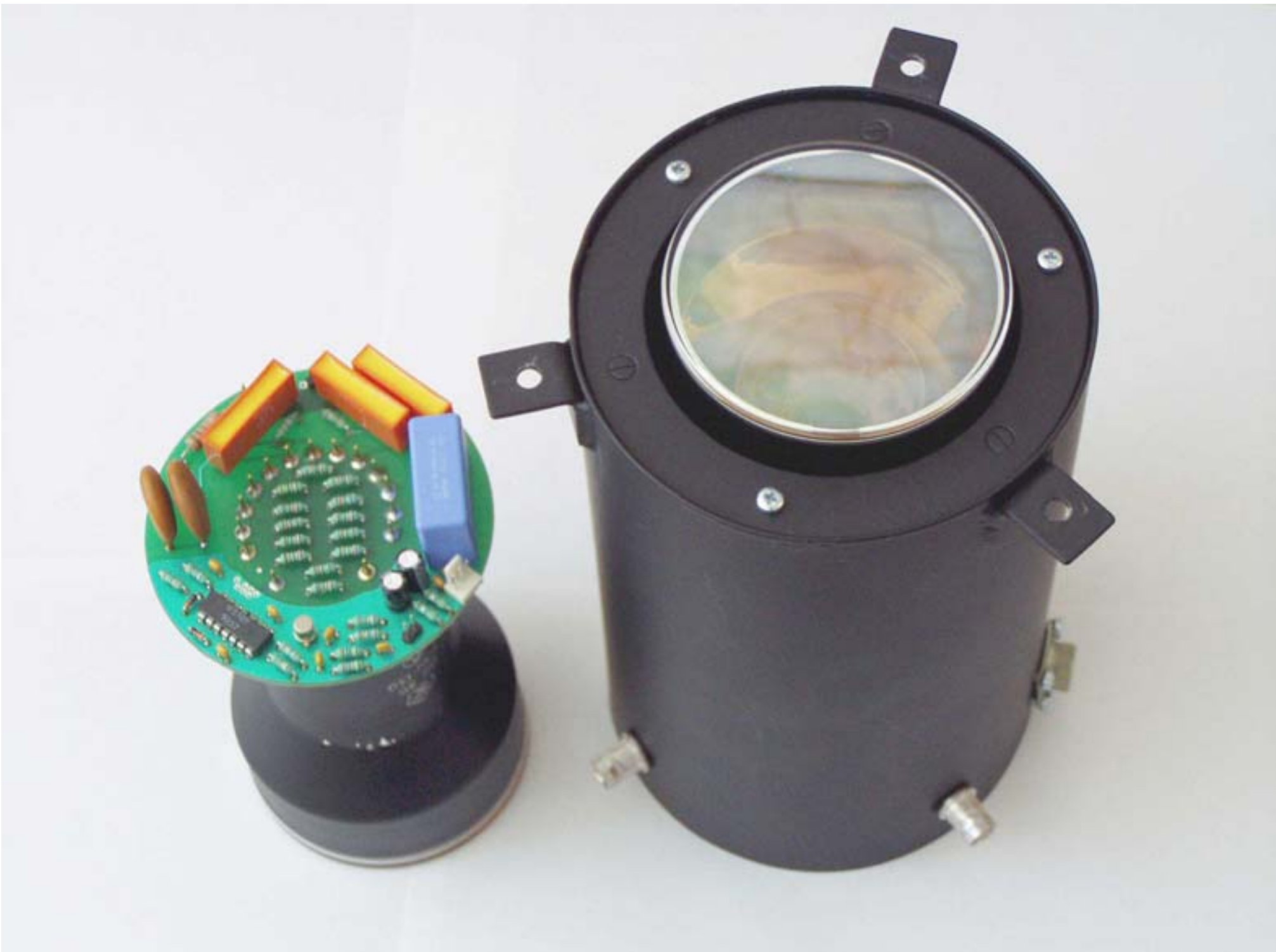

*pic. 2. A photomultiplier tube with voltage divider and preamplifier (left) and a photomultipliers housing (right)*

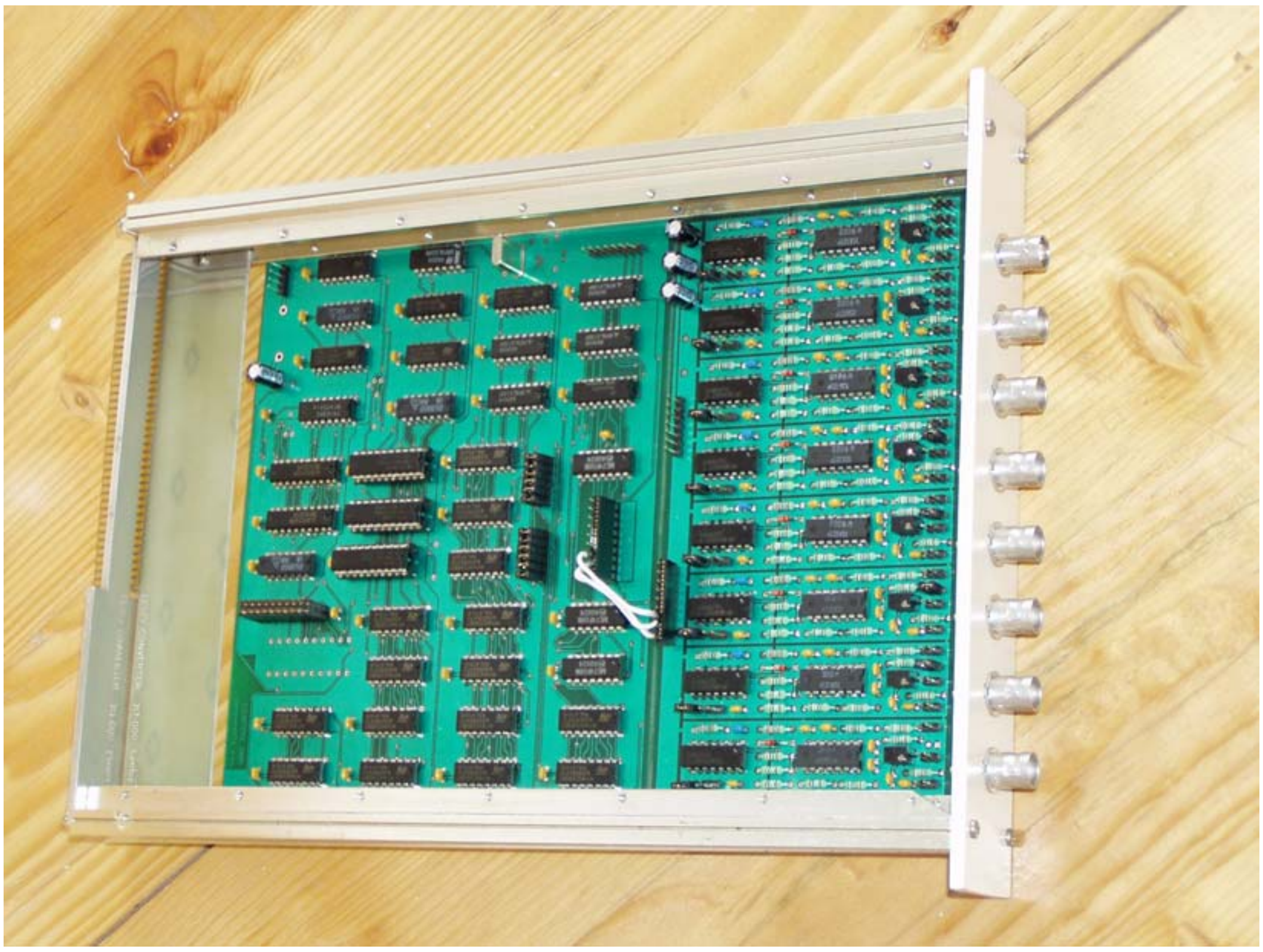

*pic. 3. The discriminators – coincidence circuits – counters block.*

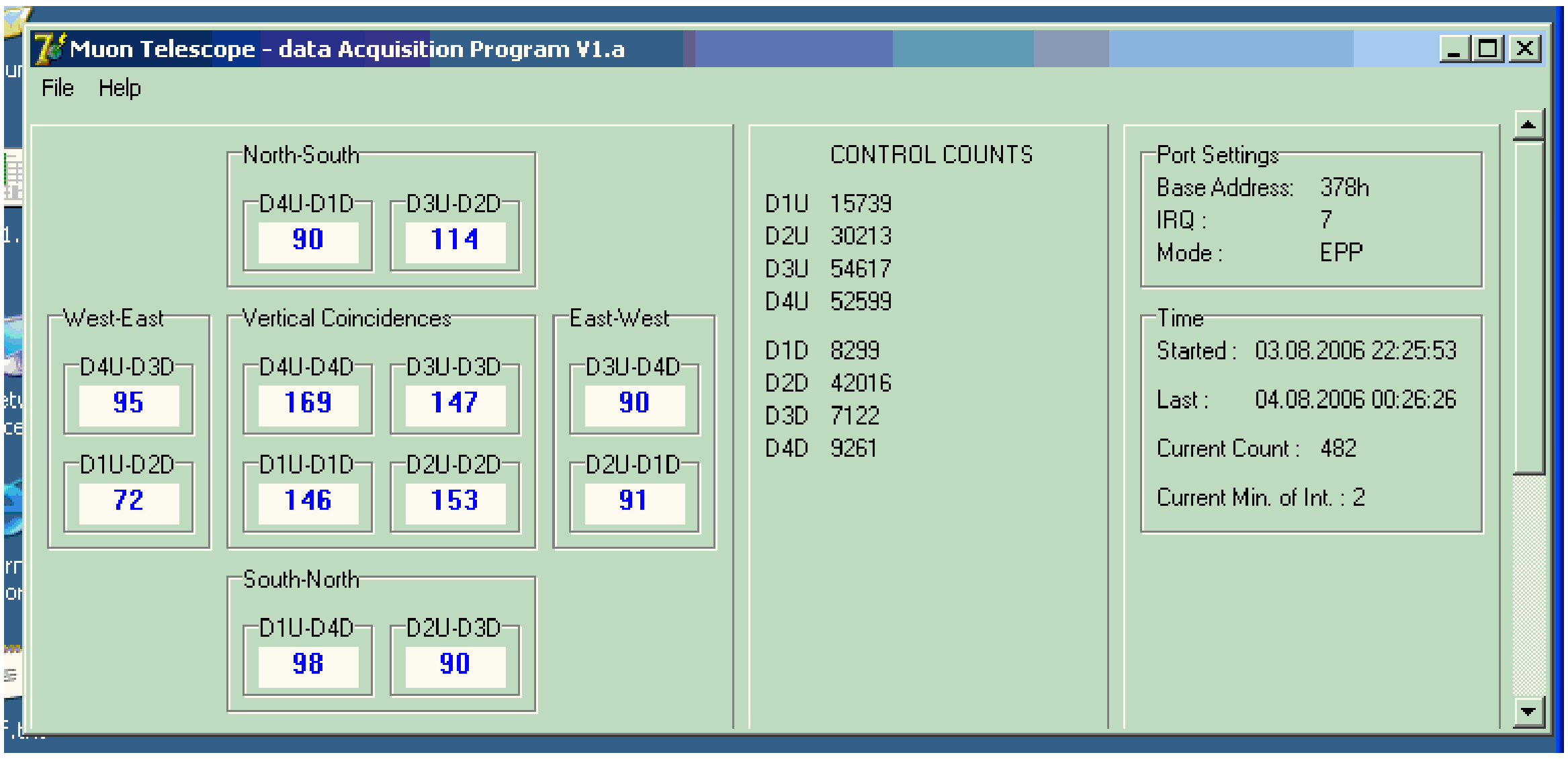

*Pic. 4. A screen shot of the data acquisition program. The number of muons passing different pairs of detectors for a 15 seconds interval is counted and can be seen on the program window.*